\documentclass{iopart}% J. Phys. B

\usepackage{iopams}
\usepackage{amssymb}
\usepackage{bm}
\usepackage{afterpage}
\usepackage{graphicx}

\def\beq{\begin{equation}}
\def\eeq{\end{equation}}
\def\nn{\nonumber}
\def\om{\omega}
\def\a{\alpha}

\def\g{\gamma}

\def\s{\sigma}
\def\D{\Delta}
\def\d{\delta}
\def\ad{a^\dagger}
\def\rd{{\rm{d}}}
\def\vp{\varphi}
\def\p{\phi}
\def\dd{{\frac{\rd}{\rd z}}}

\def\Cc{\mathbb{C}}
\def\Zz{\mathbb{Z}}
\def\Rr{\mathbb{R}}
\def\Zpl{\rm{I\!N}}

\def\Jz{\hat{J}_z}
\def\Jx{\hat{J}_x}
\def\R{\hat{R}}
\def\bal{\bar{\a}}
\def\ba{\bar{a}}
\def\bvp{\bm{\vp}}
\def\P{\hat{P}}
\def\bp{\bar{\phi}}
\def\bps{\bar{\psi}}
\def\T{\hat{T}}
\def\Hp{{\cal H}_+}
\def\Hm{{\cal H}_-}
\def\Hpm{{\cal H}_\pm}
\def\B{{\cal B}}

\begin{document}

  \title{Solution of the Dicke model for $N=3$.  }

  \author{Daniel Braak}
  \address{EP VI and
   Center for Electronic Correlations and Magnetism,\\
Institute of Physics,
    University of Augsburg, 86135 Augsburg, Germany}
  \ead{daniel.braak@physik.uni-augsburg.de}

  \date{March 21, 2013}

  \begin{abstract}
The 
$N=3$ Dicke model couples three qubits to a single radiation mode 
via dipole interaction and constitutes the simplest quantum-optical system allowing for  Greenberger-Horne-Zeilinger states.
In contrast to the case $N=1$ (the Rabi model), it is non-integrable
if the counter-rotating terms are included. 
The spectrum is determined analytically, 
employing the singularity structure of an associated
differential equation. 
While quasi-exact eigenstates known from the Rabi model do not exist, 
a novel type of spectral degeneracy becomes possible which is 
not associated with a symmetry of the system.

  \end{abstract}

  \pacs{02.30.Hq, 03.65.Ge, 42.50.Pq}

  \maketitle

\section{Introduction}
The simplest model to describe light-matter interaction is the quantum Rabi 
model, in which a two-level system (two states of a single atom in the early 
applications)
interacts with a single mode of the radiation field \cite{rabi}. A seminal step in 
its analytical treatment has been taken by Jaynes and Cummings through 
the invention of the ``rotating-wave approximation'' (RWA), valid close to resonance 
and for coupling strengths typical for atom optics \cite{JC}.  
The ensuing model can be solved analytically in a very simple way, because the
RWA introduces a strong continuous symmetry \cite{osterloh}, rendering it 
superintegrable \cite{db}. A natural generalization of the Rabi model is the Dicke model,
in which the radiation mode couples simultaneously 
to $N$ two-level systems (qubits) \cite{dicke}. It was first studied in the limit of large $N$, because
it exhibits a phase transition to a ``super-radiant'' state for strong coupling 
\cite{hepp,wang,carm}. Although the transition cannot be observed within atom optics
\cite{rzaz}, it should be realizable within circuit QED \cite{nataf,ciuti,vuk} and its equivalent 
has been experimentally observed in a Bose-Einstein condensate coupled to an 
optical cavity \cite{baum}.   

While these developments concern the Dicke model for large $N$, applications  
to quantum information technology have renewed the interest in 
the case of small $N$ \cite{sillan,haack,alt,agar,hao}. 
The model with three qubits allows in principle the dynamical generation of
Greenberger-Horne-Zeilinger (GHZ) states \cite{ghz} which could be of importance for future applications
e.g. in quantum cryptography \cite{hao}. Possible realizations of the $N=3$ Dicke model
within circuit QED will be able to explore the strong coupling region \cite{sol1} 
where the RWA is not feasible and one has to consider the full model. The $U(1)$-symmetry
induced by the RWA is so powerful that the Dicke model\footnote{Properly 
called ``Tavis-Cummings model'' if the RWA is used.}
 becomes integrable for arbitrary $N$
\cite{tavis}, 
while the model including counter-rotating terms 
is non-integrable for all $N\ge 2$ according to the criterion introduced in 
\cite{db}. The case $N=1$ is the only one where Schweber's technique \cite{schweb, reik}
or operator methods \cite{swain,tur} are applicable, therefore we shall employ in the following the method based on analysis of the associated differential equation in 
the complex domain \cite{db}.

The Dicke model for $N=3$ is described by the Hamiltonian ($\hbar=1$),
\beq
H'_D=\om\ad a +\frac{\om_0}{2}\sum_{i=1}^3\s^z_i + \frac{g'}{\sqrt{3}}(a+\ad)\sum_{i=1}^3\s^x_i.
\label{hamd1}
\eeq
Here $\om$ denotes the frequency of the radiation mode, $a$ ($\ad$) is its annihilation (creation) operator, $\om_0$ is the energy splitting of the three qubits, described by Pauli matrices $\s_i^{z,x}$, which are coupled through a dipole term with strength $g'$ to the field.  
Because the qubits are equivalent, $H_D$ is rotationally invariant, leading to a splitting of the 
eight-dimensional spin-space into irreducible components, according to
\beq
\case{1}{2}\otimes\case{1}{2}\otimes\case{1}{2}=\case{1}{2}\oplus\case{1}{2}\oplus\case{3}{2}.
\label{clebsch-gordan}
\eeq
The $N=3$ Dicke model is equivalent to two Rabi models and a system with spin $S=3/2$.  
We shall confine ourselves in the following to the $S=3/2$ model with four-dimensional spin-space.
The Hamiltonian reads with $\om=1$, $\D=\om_0/2$ and $g=g'/\sqrt{3}$,
\beq
H_D=\ad a + 2\D\Jz + 2g(a +\ad)\Jx,
\label{hamd}
\eeq
and $\Jz$ and $\Jx$ are generators of $SU(2)$ in the spin-$3/2$ representation.
$H_D$ possesses a $\Zz_2$-symmetry (parity), $\P=e^{i\pi\ad a}\otimes\R$, with
the involution $\R$ acting in spin space as $\R\Jz\R=\Jz$, $\R\Jx\R=-\Jx$.
We have $\P H_D\P=H_D$. The total Hilbert space $L^2(\Rr)\otimes\Cc^4$ splits into two invariant
subspaces (parity chains) labeled by the eigenvalues $\pm 1$ of $\P$ \cite{sol1}. 
This discrete symmetry is familiar from the Rabi model and renders it integrable because $\P$ has as many irreducible representations as the dimension of the state space of the (single) qubit \cite{db}.
The same symmetry is present in the $S=3/2$ Dicke model. However, it does not lead  to 
integrability because a spin-$3/2$ has a four-dimensional state space and the eigenstates cannot be labeled by a quantum number representing the continuous degree of freedom (the radiation field) together with a second label for the discrete degree of freedom which would have to take four different values. The $\Zz_2$-symmetry, providing only two different labels, is therefore too weak to make the $S=3/2$ Dicke model integrable. 
Nevertheless, the  symmetry leads to a 
considerable simplification of the analytical solution.

\section{The spectrum}
\begin{figure}[t!]
    \centering
    \includegraphics[width=0.7\textwidth]{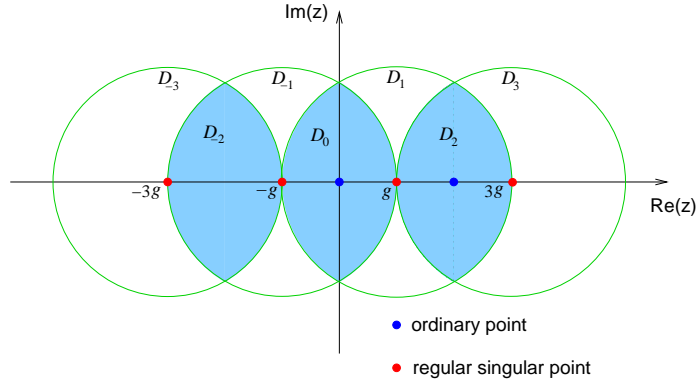}
%    \vspace*{-0.3cm}
   \caption{The singularity structure of system (\ref{sys4-a})--(\ref{sys4-d})
in the complex plane. Solutions in powers of $(z-z_j)$ converge in the disks
$D_j$ with radius $2g$ around the regular singular points $z_j=jg$.  
The two ordinary points used to define the analytic continuations are
$z_0'=0$ and $z_0=2g$.
   }
    \label{fig1}
  \end{figure}

The operator $\R$ becomes a simple reflection after transformation of $H_D$ into
``spin-boson'' form, using a  $SO(4)$-transformation $\hat{O}H_D\hat{O}=H_{sb}$, with
\beq
H_{sb}=\ad a + 2\D\Jx + 2g(a +\ad)\Jz.
\label{ham}
\eeq
As in the Rabi model, the parity invariance can be used to partially diagonalize $H_{sb}$. Define $\T=e^{i\pi\ad a}$ and 
\beq
U=\frac{1}{\sqrt{2}}\left(\begin{array}{cccc}
1&0&1&0\\
0&1&0&1\\
0&\T&0&-\T\\
\T&0&-\T&0
\end{array}
\right).
\eeq
Then $U^\dagger H_{sb}U=H_++H_-$, where $H_\pm$ acts in $\Hpm$; $\Hp$ and $\Hm$ are the two mutually orthogonal subspaces with fixed parity.
We have
\beq
H_\pm=\ad a 
+\D
\left(
\begin{array}{cc}
0&\sqrt{3}\\
\sqrt{3}&\pm2\T
\end{array}
\right)
-g(a + \ad)
\left(
\begin{array}{cc}
3&0\\
0&1
\end{array}
\right).
\label{hapm}
\eeq
We shall now represent the continuous degree of freedom in the Bargmann space
$\B$, spanned by analytic functions $f(z)$ \cite{barg}; $\Hpm$ is 
isomorphic to $\B\otimes\Cc^2$. 
The operator $\T$ acts on elements of $\B$ as $(\T f)(z)=f(-z)$.
The eigenvalue equation
$H_+\bvp=E\bvp$ takes with $\bvp=(\p_1(z),\p_2(z))^T$ the form 
of a non-local system of linear ordinary differential equations in the complex domain,
\begin{eqnarray}
\fl
z\dd\p_1(z) +\sqrt{3}\D\p_2(x)-3gz\p_1(z)-3g\dd\p_1(z) &=E\p_1(z),
\label{sys2-a}\\
\fl
z\dd\p_2(z)+\sqrt{3}\D\p_1(z)-gz\p_2(z)-g\dd\p_2(z)+2\D\p_2(-z) &=E\p_2(z).
\label{sys2-b}
\end{eqnarray}
With the definitions $\bp_j(z)=\p_j(-z)$, $j=1,2$ and denoting the derivative with a prime, we obtain the following
local system of the first order,
\begin{eqnarray}
(z-3g)\p'_1 &= (E+3gz)\p_1-\sqrt{3}\D\p_2,
\label{sys4-a}\\
(z-g)\p'_2 &=  (E+gz)\p_2-\sqrt{3}\D\p_1-2\D\bp_2,
\label{sys4-b}\\
(z+3g)\bp'_1 &= (E-3gz)\bp_1-\sqrt{3}\D\bp_2,
\label{sys4-c}\\
(z+g)\bp'_2 &=  (E-gz)\bp_2-\sqrt{3}\D\bp_1-2\D\p_2.
\label{sys4-d}
\end{eqnarray}
The system (\ref{sys4-a})--(\ref{sys4-d}) has four regular singular points at
$z=\pm g,\pm 3g$, and an irregular singular point at 
infinity \cite{ince}. The latter
has rank 1 as in the Rabi model \cite{db2}, therefore a solution $\bvp$ of
(\ref{sys2-a},\ref{sys2-b}) is an eigenvector of $H_+$ with eigenvalue $E$
if and only if $\p_1(z)$ and $\p_2(z)$ are analytic in the 
whole complex plane \cite{barg}.  
The indicial analysis of  (\ref{sys4-a})--(\ref{sys4-d}) shows that the exponents
at the points $z=\pm 3g$ are 0 and $E+9g^2$, whereas at  $z=\pm g$ they are
0 and $E+g^2$. The exponent zero is three-fold degenerate at all regular 
singularities. This is the major difference to the Rabi model, which has 
only two regular singular points at $\pm g$ and the exponent zero is 
non-degenerate at each 
of these points.
Formal solutions of (\ref{sys4-a})--(\ref{sys4-d}) in terms of power series
in $z-z_j$ are possible in regions $D_j$ with $z_j=jg$, $j=\pm1,\pm3$, 
see Fig.~\ref{fig1}.
%%%%%%%%%%%%%%%%%%
%
%   fig 1: regions in the z-plane
%
%%%%%%%%%%%%%%%%%%
The expansion around $z_1=g$ reads for $j=1,2$,
\beq
\p_j(z)=\sum_{n=0}^\infty\a_{j,n}(z-g)^n,\quad
\bp_j(z)=\sum_{n=0}^\infty\bal_{j,n}(z-g)^n.
\label{exp-phi}
\eeq
The expansion (\ref{exp-phi}) is absolutely convergent in $D_1$, 
with radius of convergence $2g$ and $\p_j(z)$, $\bp_j(z)$ are analytic at $z_1$. 
Likewise, there is an expansion of $\p_j(z),\bp_j(z)$ around
$z_3$, which we denote as $\psi_j(z), \bps_j(z)$,
\beq
\psi_j(z)=\sum_{n=0}^\infty a_{j,n}(z-3g)^n,\quad
\bps_j(z)=\sum_{n=0}^\infty\ba_{j,n}(z-3g)^n.
\label{exp-psi}
\eeq
(\ref{exp-psi}) converges in $D_3$ and $\psi_j(z)$, $\bps_j(z)$ are 
analytic at $z_3$. Using the identification $\bp_j(z)=\p_j(-z)$ and
$\bps_j(z)=\psi_j(-z)$, (\ref{exp-phi}) and (\ref{exp-psi}) furnish
series expansions of $\p_j(z)$ in regions $D_{-1}$ and $D_{-3}$, respectively.
These series lead for arbitrary $E$ to functions which are 
analytic at their expansion points but develop branch-cuts at the other singular points. The
discrete set of eigenvalues $E_n$, $n=0,1,2,\ldots$ is 
determined by the condition that all four expansions describe the 
same function $\p_j(z)$, i.e. that they are analytic continuations of each other \cite{db2}.

Inserting the ansatz (\ref{exp-phi}) into (\ref{sys4-a})--(\ref{sys4-d}) 
yields the following coupled recurrence relations for $n\ge 0$ and with 
$x=E+g^2$,
\begin{eqnarray}
\fl
2g(n+1)\big(x-(n+1)\big)\a_{2,n+1} =
(3\D^2-n^2+2xn-2g^2-x^2-2xg^2)\a_{2,n} \nn \\
%\fl
  + g(1-4x-2g^2+4n-4)\a_{2,n-1} 
 - 3g^2\a_{2,n-2} \label{rec-phi-a} \\
%\fl
  + 4\D g(n+1)\bal_{2,n+1}
+2\D(x+2g^2-n)\bal_{2,n}
+6\D g\bal_{2,n-1}, \nn
\end{eqnarray}
and
\begin{eqnarray}
\fl
8g^2(n+2)(n+1)\bal_{2,n+2}=
-2g(n+1)(3n+2-3x+8g^2)\bal_{2,n+1} \nn \\
%\fl
+\big(3\D^2-n^2+n(2x-16g^2)-4g^2-(x-2g^2)(x-4g^2)\big)\bal_{2,n} \nn \\
%\fl
+(4x-10g^2+3-4n)\bal_{2,n-1} 
-3g^2\bal_{2,n-2} \label{rec-phi-b}\\
%\fl
-8\D g(n+1)\a_{2,n+1}
+2\D(x-4g^2-n)\a_{2,n}
-6\D g\a_{2,n-1}. \nn
\end{eqnarray}
The coefficients $\a_{1,n},\bal_{1,n}$ read in terms of the $\a_{2,n},\bal_{2,n}$ as
\begin{eqnarray}
\fl
\a_{1,n} &=& -\frac{1}{\sqrt{3}\D}[(n-x)\a_{2,n}-g\a_{2,n-1}+2\D\bal_{2,n}],
\label{phi1}\\
\fl
\bal_{1,n} &=& -\frac{1}{\sqrt{3}\D}[(n-x+2g^2)\bal_{2,n}+g\bal_{2,n-1}
+2g(n+1)\bal_{2,n+1}+2\D\a_{2,n}].
\label{phi1b}
\end{eqnarray}
The system (\ref{rec-phi-a},\ref{rec-phi-b}) cannot be reduced to a linear three-term recurrence relation, therefore continued-fraction techniques are not applicable to the present model \cite{moroz}. Moreover, the relations 
(\ref{rec-phi-a})--(\ref{phi1b}) are not sufficient to determine the eigenfunctions
uniquely because there are three linear independent solutions of 
(\ref{rec-phi-a},\ref{rec-phi-b}) which are analytic at $z_1$. The initial conditions for (\ref{rec-phi-a},\ref{rec-phi-b}) are given by the set
$\{\a_{2,0},\bal_{2,0},\bal_{2,1}\}$ and the three solutions 
$\{\p^{(k)}_j(z),\bp_j^{(k)}(z)\}$ of 
(\ref{sys4-a})--(\ref{sys4-d}) analytic at $z=g$ are obtained by setting one element of the
set to 1 and the others to 0.

The general solution of this type is therefore given as
\beq
\p_j(z;\{\g_k\})=\sum_{k=1}^3\g_k\p_j^{(k)}(z),\qquad
\bp_j(z;\{\g_k\})=\sum_{k=1}^3\g_k\bp_j^{(k)}(z),
\label{genphi}
\eeq
with $\g_k$ to be determined. In the same way, we obtain for
the $\psi_j(z),\bps_j(z)$ the recurrences,
\begin{eqnarray}
\fl
2g(n+1)(n-x-8g^2)a_{2,n+1} =
(3\D^2-n^2+n(2x+16g^2)-(x+8g^2)(x+2g^2))a_{2,n} \nn \\
%\fl
  + g(4n-3-4x-14g^2)a_{2,n-1} 
 - 3g^2a_{2,n-2} \label{rec-psi-a} \\
%\fl
+2\D(x+8g^2-n)\ba_{2,n}
+6\D g\ba_{2,n-1}, \nn
\end{eqnarray}
and
\begin{eqnarray}
\fl
24g^2(n+2)(n+1)\ba_{2,n+2}=
2g(n+1)(5x-5n-3 -32g^2)\ba_{2,n+1} \nn \\
%\fl
+\big(3\D^2-n^2+n(2x-32g^2)-6g^2-(x-10g^2)(x-4g^2)\big)\ba_{2,n} \nn \\
%\fl
+g(4x-4n+3-22g^2)\ba_{2,n-1} 
-3g^2\ba_{2,n-2} \label{rec-psi-b}\\
%\fl
-12\D g(n+1)a_{2,n+1}
+2\D(x-10g^2-n)a_{2,n}
-6\D ga_{2,n-1}. \nn
\end{eqnarray}
For the  $a_{1,n},\ba_{1,n}$ we have,
\begin{eqnarray}
\fl
a_{1,n} &=& -\frac{1}{\sqrt{3}\D}[(n-x-2g^2)a_{2,n}-ga_{2,n-1} +2g(n+1)a_{2,n+1}+2\D\ba_{2,n}],
\label{psi1}\\
\fl
\ba_{1,n} &=& -\frac{1}{\sqrt{3}\D}[(n-x+4g^2)\ba_{2,n}+g\ba_{2,n-1}
+4g(n+1)\ba_{2,n+1}+2\D a_{2,n}].
\label{psi1b}
\end{eqnarray}
The solution space of (\ref{sys4-a})--(\ref{sys4-d}) analytic at $z=3g$ is likewise 
three-dimensional, determined by the initial set $\{a_{2,0},\ba_{2,0},\ba_{2,1}\}$
and the general solution reads 
\beq
\psi_j(z;\{c_k\})=\sum_{k=1}^3c_k\psi_j^{(k)}(z),\qquad
\bps_j(z;\{c_k\})=\sum_{k=1}^3c_k\bps_j^{(k)}(z),
\label{genpsi}
\eeq
with three unknown constants $c_k$. (\ref{genphi}) and (\ref{genpsi}) are the 
most general solutions analytic at $g$ and $3g$ respectively, if the spectral parameter $x$ is not a
positive integer (see (\ref{rec-phi-a})) or satisfies $x+8g^2=n$, $n=0,1,2,\ldots$ (\ref{rec-psi-a}). 
These special values for $x$ determine the two types of baselines in 
the model where the exceptional spectrum is located (the Rabi model has only one type of baseline). 
We shall now determine the condition under which all four sets of series expansions 
for $\p_1(z),\p_2(z)$ describe the same functions, which are therefore analytic in the
 whole complex plane and correspond to an eigenvector $(\p_1,\p_2)^T$ of $H_+$.  

\begin{figure}[t!]
    \centering
    \includegraphics[width=0.7\textwidth]{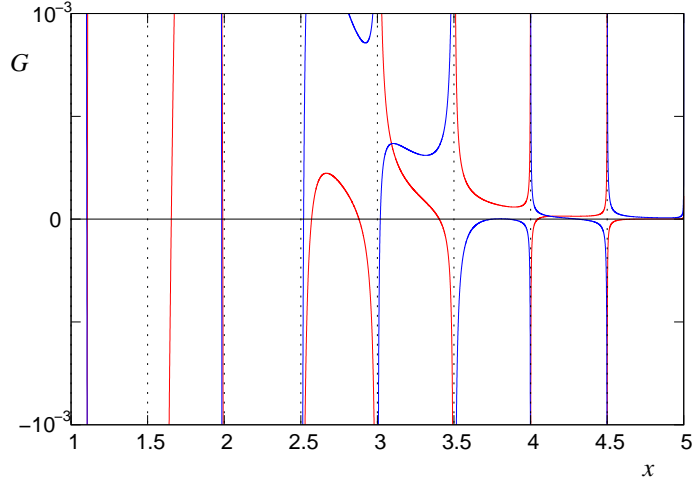}
%    \vspace*{-0.3cm}
   \caption{The $G$-functions of the Dicke model for positive (red) and negative (blue) parity.
$\D=0.7$ and $g=0.25$, therefore the baselines of 
first kind are located at integer, and the baselines of second kind at half-integer values. At both kinds
has 
$G^D_\pm(x,0.5)$ poles of order 1 or 3. 
   }
    \label{fig2}
  \end{figure}

Because the functions $\{\p_1(z),\p_2(z),\bp_1(z),\bp_2(z)\}$ 
satisfy the same differential equation of the first
order as $\{\psi_1(z),\psi_2(z),\bps_1(z),\bps_2(z)\}$, both sets will coincide in
$D_2=D_1\cap D_3$, if they coincide at one regular point $z_0\in D_2$ \cite{db2}.
This yields four equations for the functions in (\ref{genphi}) and (\ref{genpsi}).
Furthermore,  $\{\p_1(z),\p_2(z),\bp_1(z),\bp_2(z)\}$ and 
$\{\bp_1(-z),\bp_2(-z),\p_1(-z),\p_2(-z)\}$ satisfy the same differential equation
and coincide in  all of $D_0=D_1\cap D_{-1}$ if they do so at a point $z_0'\in D_0$. 
Obviously, only two of the four equations are independent if $z_0'=0$. 
$(\bp_1(-z),\bp_2(-z))^T$ is then the analytic continuation of $(\p_1(z),\p_2(z))^T$
into the disk $D_{-1}$. But because $\bp_j(z)=\bps_j(z)$ for $z\in D_2$,
it follows that $\bps_j(-z)$ is the analytic continuation of $\bp_j(-z)$ 
(and therefore of $\p_j(z)$) into the disk $D_{-3}$. The six equations
\beq
\p_j(z_0)=\psi_j(z_0), \quad \bp_j(z_0)=\bps_j(z_0), \quad \p_j(0)=\bp_j(0), 
\label{cond1}
\eeq
for $j=1,2$ and $z_0\in D_2$ are  equivalent to the analyticity of
$\p_j(z)$ in $\Cc$. A non-trivial solution of (\ref{cond1}) can be found if the parameters $\{\g_k,c_k\}$ are not all zero. 
The functions $\p_j(z),\bp_j(z),\ldots$ depend parametrically on the energy $E=x-g^2$.
Define then the matrix $M_+(x,z_0)$  as
\beq
\fl
M_+=
\left(
\begin{array}{cccccc}
\psi_1^{(1)}(z_0) & \psi_1^{(2)}(z_0) & \psi_1^{(3)}(z_0) & -\p_1^{(1)}(z_0) & -\p_1^{(2)}(z_0) & -\p_1^{(3)}(z_0)\\
\psi_2^{(1)}(z_0) & \psi_2^{(2)}(z_0) & \psi_2^{(3)}(z_0) & -\p_2^{(1)}(z_0) & -\p_2^{(2)}(z_0) & -\p_2^{(3)}(z_0)\\
\bps_1^{(1)}(z_0) & \bps_1^{(2)}(z_0) & \bps_1^{(3)}(z_0) & -\bp_1^{(1)}(z_0) & -\bp_1^{(2)}(z_0) & -\bp_1^{(3)}(z_0)\\
\bps_2^{(1)}(z_0) & \bps_2^{(2)}(z_0) & \bps_2^{(3)}(z_0) & -\bp_2^{(1)}(z_0) & -\bp_2^{(2)}(z_0) & -\bp_2^{(3)}(z_0)\\
 0&0& 0 &  \d\p^{(1)}_1(0) &  \d\p^{(2)}_1(0) &   \d\p^{(3)}_1(0)\\ 
 0&0& 0 &  \d\p^{(1)}_2(0) &  \d\p^{(2)}_2(0) &   \d\p^{(3)}_2(0)\\ 
\end{array}
\right),
\label{matrix}
\eeq
with $\d\p_j^{(k)}(0)=\bp_j^{(k)}(0)-\p_j^{(k)}(0)$. It follows that the $G$-function of the Dicke model
for positive parity,
\beq
G^D_+(x,z_0)=\det M_+(x,z_0),
\label{G-dicke}
\eeq 
is zero for all $z_0\in D_2$, if and only if $x=E+g^2$ corresponds 
to an element of the spectrum of $H_+$. The discrete set of zeros $x_n$ with $G^D_+(x_n,z_0)=0$ for $n=0,1,2,\ldots$ 
determines the
{\it regular} spectrum $\s_r(H_+)= \{x_n-g^2\}_{n\in\Zpl_0}$ of $H_+$ \cite{db}. 
The regular spectrum of $H_-$ is given in an analogous manner, starting from recurrences 
(\ref{rec-phi-a},\ref{rec-phi-b},\ref{rec-psi-a},\ref{rec-psi-b}) by reversing the sign of 
the coupling terms
between $\a_n$ and $\bal_n$, resp. $a_n$ and $\ba_n$ (see (\ref{hapm})), constructing functions
$\p_{j,-}^{(k)}(x,z)$, $\bp_{j,-}^{(k)}(x,z),\ldots$ and the matrix $M_-(x,z_0)$.
Fig.~\ref{fig2} 
shows $G^D_\pm(x,2g)$ for $g=0.25$ and $\D=0.7$. 
%%%%%%%%%%%%%%%%%%%%%%%%%
%
%   
%   fig 2: G-functions
%
%
%%%%%%%%%%%%%%%%%%%%%%%%%

The functions 
$\p_{j,-}^{(k)}(x,z)$, $\bp_{j,-}^{(k)}(x,z)$ entering $M_-(x,z_0)$ 
are {\it not} related in a simple manner to their
counterparts $\p_{j,+}^{(k)}(x,z)$, $\bp_{j,+}^{(k)}(x,z)$ for positive parity, 
in contrast to the case of the Rabi model. There one has only one pair of functions 
$\{\p_\pm(x,z)$, $\bp_\pm(x,z)\}$
and the $G$-function of the Rabi model reads simply
\beq
G^R_\pm(x)=\bp_\pm(x,0)-\p_\pm(x,0).
\label{G-rabi}
\eeq
Moreover, $\p_-(x,z)=\p_+(x,z)$, $\bp_-(x,z)=-\bp_+(x,z)$. It follows at once that the regular spectrum of
the Rabi model is not degenerate between states of different parity, as $G^R_+(x)=G^R_-(x)=0$ implies
$\p_\pm(x,z)\equiv 0$. On the other hand, regular states with the same parity cannot be degenerate because
the formal solution $\p_+(x,z)$, analytic at $z=g$, is unique for $x\notin\Zpl$. The only possibility 
for degenerate eigenvalues in the Rabi model occurs therefore in the {\it exceptional} spectrum, where $x\in \Zpl$.  The two degenerate states
at these points have different parity because there are only two
linear independent formal solutions of the eigenvalue equation
and the pole of $G^R_\pm(x)$ at integer $x$ is lifted in both $G^R_+(x)$ and 
$G^R_-(x)$, which corresponds to a condition satisfied by the 
model parameters $g$ and $\D$ \cite{kus-lew,db}.

\begin{figure}[t!]
    \centering
    \includegraphics[width=0.7\textwidth]{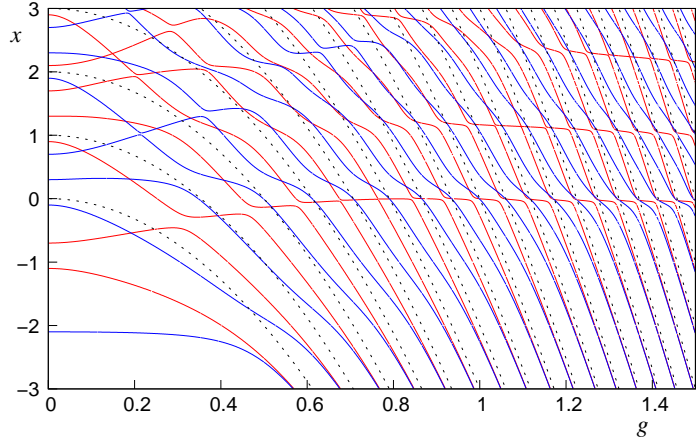}
%    \vspace*{-0.3cm}
   \caption{The spectrum of the Dicke model for even (red) and odd (blue) parity at $\D=0.7$ and 
for varying $g$. The $y$-axis shows $x=E+g^2$, baselines of first kind are horizontal straight lines. 
The ground state has odd parity as in the Rabi model. The two ladders of eigenvalues
with different parity intersect within the regular spectrum. There are no degeneracies (but narrow avoided crossings) for fixed parity in this 
parameter window. The baselines of first kind (not depicted) and of second kind (dashed lines)
emerge as limiting values in the deep strong coupling regime $g>1$.
   }
    \label{fig3}
  \end{figure}

\begin{figure}[t!]
    \centering
    \includegraphics[width=0.7\textwidth]{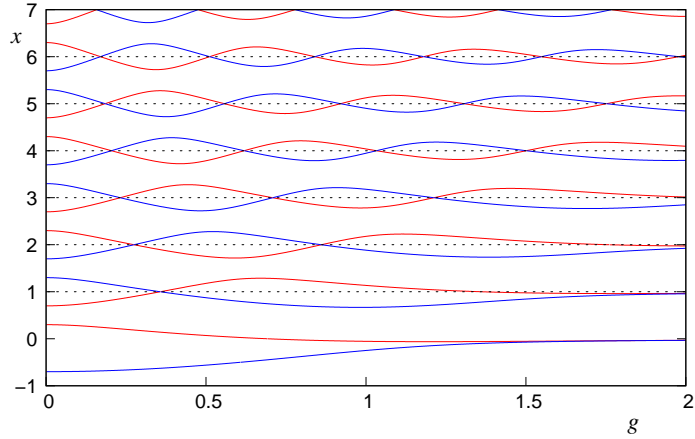}
%    \vspace*{-0.3cm}
   \caption{The Rabi spectrum for the same parameters as in Fig.~\ref{fig3}. Degeneracies occur 
solely between states of different parity and are always located on the baselines. 
   }
    \label{fig4}
  \end{figure}

As mentioned above, $G^D_\pm(x,z_0)$ has two different types of baselines, located at
$x=n, n\in \Zpl$ (first kind) and $x=n-8g^2, n\in \Zpl_0$ (second kind). 
Although $G^D_\pm(x,z_0)$ has pole singularities in $x$
at these values for general $g,\D$, there may be (exceptional) eigenvalues $E_e^1=n-g^2$, resp.
$E_e^2=n-9g^2$, if the singularity is lifted in $G^D_+(x,z_0)$ or  $G^D_-(x,z_0)$.
This exceptional solution, however, is usually non-degenerate, because there is no single lifting condition
(as in the Rabi model), valid for {\it both} parities.\footnote{
Non-degenerate exceptional solutions appear in the Rabi model as well, but then the pole is not lifted
in $\p_\pm(x,0)$ but only in the difference, either $\bp_+(x,0)-\p_+(x,0)$ or $\bp_-(x,0)-\p_-(x,0)$.
These solutions are exceptional but not quasi-exact.}
Therefore, a quasi-exact spectrum in the sense of the Rabi model is not present in the Dicke model.
If one defines the quasi-exact spectrum differently, by demanding that the eigenfunctions 
are  polynomial in $z$ (apart from a common factor), this possibility is not ruled out in 
principle, although
the set $\{n,g,\D\}$ would then have to satisfy three consistency equations \cite{kus-lew}, 
making a solution with integer $n$ unlikely. In fact, as these solutions lie necessarily on baselines and 
are parity degenerate, the consistency equations given by K\'us and Lewenstein must comprise the
two independent lifting conditions for $G^D_\pm(x,z_0)$. 
On the other hand, $G^D_+(x,z_0)=G^D_-(x,z_0)$ is no longer tantamount to  vanishing of the
wave-functions $\p_{j,\pm}(x,z)$ themselves, therefore regular eigenvalues of the Dicke model may 
well be parity degenerate. Figs.~\ref{fig3} and \ref{fig4} show the Dicke and Rabi spectra for fixed
$\D$ and varying $g$. It is apparent that the coupling between exceptional eigenvalues       
and degeneracies renders the Rabi spectrum much more ``regular'' than the Dicke spectrum, apart from
the complication coming from two kinds of baselines in the latter. 
%%%%%%%%%%%%%%%%%%%%%%%%%
%
%   
%   fig 3: Dicke spectrum
%
%
%%%%%%%%%%%%%%%%%%%%%%%%%
%%%%%%%%%%%%%%%%%%%%%%%%%
%
%   
%   fig 4: Rabi spectrum
%
%
%%%%%%%%%%%%%%%%%%%%%%%%% 
  
All regular eigenvalues of the Rabi model with fixed parity correspond to unique eigenfunctions because
the exponent 0 of the indicial equation at each singular point is 
non-degenerate \cite{ince,kus-lew} and there exists at most one solution analytic 
at $z=\pm g$. In contrast, we have three  solutions analytic at each of the $z_j$ in 
the case of the $N=3$ model. Although generally there is only one 
solution analytic at all four singular points, the possibility is not excluded that the kernel
of $M_\pm(x,z_0;g,\D)$ has dimension $>1$ at some value $x$, which would correspond to a degeneracy {\it within} a given parity chain $\Hpm$. 
Indeed, for this it is necessary that the linear term in the characteristic polynomial of $M_\pm(x,z_0;g,\D)$ vanishes, providing a condition to be satisfied by the model parameters $g$ and $\D$, in analogy to the equation determining the quasi-exact spectrum of the Rabi model \cite{kus-lew}.
On has to note here that this equation is not independent from $\det M_\pm(x,z_0;g,\D)=0$, 
because the value of $x$ is not restricted to integers. Both equations form a coupled system
to determine the triple $(x_{\rm deg},g_{\rm deg},\D_{\rm deg})$ where two states with equal parity are degenerate.  
If such a triple exists, a level crossing at $E=x_{\rm deg}-g^2_{\rm deg}$ will
appear within the corresponding parity 
chain $\Hpm$.
These degeneracies would not originate from a global symmetry of the model
and thus are  not accidental, because the degenerate states do not belong to dynamically decoupled
subspaces. The numerical investigations done so far have shown no hint to this novel type of degeneracy yet. However, in a recent work on the $N=2$ model with inequivalent qubits,
Chilingaryan and Rodr\'iguez-Lara have discovered 
level crossings within spectra with fixed parity \cite{chil}. This is a strong 
indication that the phenomenon predicted here for the $N=3$ model is not forbidden 
by some special feature of the matrix $M_\pm(x,z_0;g,\D)$.
 
\section{Conclusions}
We have computed the spectrum of the Dicke model for three qubits analytically using the technique based on formal solutions in the Bargmann space, where the 
spectral condition corresponds to 
analyticity in the whole complex plane \cite{barg,schweb,db}. In contrast to the
$N=1$ model, the argument used by Schweber to derive a continued-fraction
representation of the spectral condition is not applicable, because the
formal solutions are not given in terms of linear three-term
recurrence relations \cite{moroz}. If one defines the model on a truncated
Hilbert space, matrix-valued continued fractions could be employed 
in principle \cite{durst}, but this approach 
suffers from ambiguities and can be justified 
only in the case $N=1$, by its formal equivalence to 
Schwebers method \cite{db3}. 

The Dicke model shows much more spectral ``irregularities'' 
than the Rabi model, whose spectral graph is restricted by the position of the quasi-exact eigenvalues, confining degeneracies to the baselines. The $N=3$ model possesses two types of baselines, but they govern only the asymptotics for strong coupling (Fig.~\ref{fig3}); the quasi-exact spectrum does not exist and levels corresponding to
different parity intersect within the regular spectrum, while the exceptional spectrum is non-degenerate. On the other hand, the structure of the $G$-function for $N=3$ predicts degeneracies {\it within} the parity chains, which are forbidden for $N=1$. These degeneracies are not related to a
symmetry of the model and do not imply integrability. 
They cannot be termed ``accidental'' either, as the degenerate states do not belong to different invariant subspaces.
However, if they appear in sufficient number, which seems to be possible for large $N$, it could lead to a level statistics resembling Poissonian behavior, which has been found numerically for $N\ge 20$ in the coupling region below the quantum phase transition \cite{emary}. The implications of this novel type of degeneracy for the notion of (non)-integrability in systems with less than two continuous degrees of freedom \cite{db} have yet to be explored.

 \ack
This work was supported by Deutsche Forschungsgemeinschaft through TRR~80.


\section{References}

  \begin{thebibliography}{99}



  \bibitem{rabi}  Rabi I I 1936  {\it Phys. Rev.} 
                  {\bf 49} 324


\bibitem{JC}  Jaynes E T and Cummings F W  1963 {\it Proc. IEEE}
              {\bf 51} 89

\bibitem{osterloh} Amico L, Frahm H, Osterloh A and Ribeiro G A P 2007 {\it Nucl. Phys. B}
                   {\bf 787} 283
 
\bibitem{db}  Braak D 2011 {\it Phys. Rev. Lett.}
              {\bf 107} 100401 

\bibitem{dicke} Dicke R H 1954 {\it Phys. Rev.} 
                {\bf 93} 99

\bibitem{hepp}  Hepp K and Lieb E H  1973 {\it Ann. Phys. NY}
                {\bf 76} 360 

\bibitem{wang}  Wang Y K and Hioe F T 1973 {\it Phys. Rev. A}
                {\bf 7} 831

\bibitem{carm} Carmichael H J, Gardiner C W and Walls D F 1973 {\it Phys. Lett. A}
                {\bf 46} 47 

\bibitem{Nahmad} Nahmad-Achar E, Casta\~nos O, L\'opez-Pe\~na R and Hirsch J G 2013 {Phys. Scr.}
                {\bf 87} 038114 

\bibitem{rzaz} Rza\.zewski K, W\'odkiewicz K and \.Zakowicz W 1975 {\it Phys. Rev. Lett.}
               {\bf 35} 432

\bibitem{nataf} Nataf P and Ciuti C 2010 {\it Nature Comm.}
                {\bf 1} 72

\bibitem{ciuti} Nataf P and Ciuti C 2010 {\it Phys. Rev. Lett.}
                {\bf 104} 023601 

\bibitem{vuk}  Vukics A and Domokos P 2012 {\it Phys. Rev. A}
                {\bf 86} 053807

\bibitem{baum}  Baumann K, Guerlin C, Brennecke F and Esslinger T 2010 {\it Nature}
                {\bf 464} 1301 

\bibitem{sillan}  Sillanp\"a\"a M A, Park J I and Simmons R W 2007 {\it Nature}
                  {\bf 449} 438                

\bibitem{haack}  Haack G, Helmer F, Mariantoni M, Marquardt F and Solano E 2010 {\it Phys. Rev. B}
                 {\bf 82} 024514
 
\bibitem{alt}   Altintas F and Eryigit R 2012 {\it Phys. Lett. A}
                {\bf 376} 1791

\bibitem{agar}  Agarwal S, Hashemi Rafsanjani S M and Eberly J H 2012 {\it Phys. Rev. A} 
                {\bf 85} 043815

\bibitem{hao}   Hao S Y, Xia Y, Song J and An N B 2013 
                arXiv:1301.0684

\bibitem{ghz}  Greenberger D M, Horne M, Shimony A and Zeilinger A 1990 {\it Am. J. Phys.}
               {\bf 58} 1131 

\bibitem{sol1} Casanova J, Romero G, Lizuain I, Garc\'ia-Ripoll J J and Solano E 2010 
               {\it Phys. Rev. Lett.}
               {\bf 105} 263603

\bibitem{tavis} Tavis M and Cummings F W 1968 {\it Phys. Rev.}
               {\bf 170} 379 

\bibitem{schweb} Schweber S 1967 {\it Ann. Phys., NY}
                 {\bf 41} 205

\bibitem{reik}  Reik H G, Nusser H and Amarante Ribeiro L A 1982 {\it J. Phys. A: Math. Gen.} 
                {\bf 15} 3491

\bibitem{swain} Swain S 1973 J. Phys. A: Math. Nucl. Gen. 
                {\bf 6} 192

\bibitem{tur}   Tur E A 2000 {\it Opt. Spectrosc.}
                {\bf 89} 574 


\bibitem{barg} Bargmann V 1961 {\it Commun. Pure Appl. Math.}
               {\bf 14} 187

\bibitem{ince} Ince E L 1956 {\it Ordinary Differential Equations} (New York: Dover)

\bibitem{db2} Braak D 2013 {\it Ann. Phys. (Berlin)}
              {\bf 525} L23

\bibitem{moroz}  Moroz A 2012 {\it Europhys. Lett.} 
                 {\bf 100} 60010

\bibitem{kus-lew} Ku\`s M and Lewenstein M 1986 {\it J. Phys A: Math. Gen.}
                 {\bf 19} 305

\bibitem{chil} Chilingaryan S A and Rodr\'iguez-Lara B M 2013 
                arXiv:1301.4462 

\bibitem{durst} Durst C, Sigmund E, Reineker P and Scheuing A 1986 
                {\it J. Phys. C: Solid State Phys.}
                {\bf 19} 2701
\bibitem{db3}   Braak D 2012
                arXiv:1205.3439 (to appear in {\it J. Phys. A: Math. Theor.}) 

\bibitem{emary} Emary C and Brandes T 2003 {\it Phys. Rev. Lett.}
                {\bf 90} 044101



  \end{thebibliography}
\end{document}